\documentclass[aps,twocolumn,pre,showpacs,groupedaddress,showkeys,amsmath,amssymb,superscriptaddress]{revtex4}

\usepackage{float}

\usepackage{amsmath}
\usepackage{amssymb}
\usepackage{graphicx}
\usepackage{bm}
\usepackage{epic}
\usepackage{eepic}
\usepackage{pifont}
\usepackage[utf8]{inputenc}
\usepackage{rotating}
\usepackage{color}
\usepackage{nicefrac}
\usepackage{ulem}
\usepackage[caption=false]{subfig}
\usepackage{array}
\usepackage{cleveref}
\begin{document}
\draft
\title{Squeezing multiple soft particles into a constriction: transition to clogging}
\author{Clément Bielinski}
\affiliation{Biomechanics and Bioengineering Laboratory (UMR CNRS 7338),\\
CNRS, Université de Technologie de Compiègne,\\
60200 Compiègne, France}
\author{Othmane Aouane}
\affiliation{Helmholtz Institute Erlangen-Nürnberg for Renewable Energy,\\
Forschungszentrum Jülich,\\
Cauerstr. 1, 91058 Erlangen, Germany}
\author{Jens Harting}
\affiliation{Helmholtz Institute Erlangen-Nürnberg for Renewable Energy,\\
Forschungszentrum Jülich,\\
Cauerstr. 1, 91058 Erlangen, Germany}
\affiliation{Department of Chemical and Biological Engineering and Department of Physics, Friedrich-Alexander-Universit\"{a}t Erlangen-N\"{u}rnberg,
Cauerstr. 1, 91058 Erlangen, Germany}
\author{Badr Kaoui}
\email{badr.kaoui@utc.fr}
\affiliation{Biomechanics and Bioengineering Laboratory (UMR CNRS 7338),\\
CNRS, Université de Technologie de Compiègne,\\
60200 Compiègne, France}
\begin{abstract}
We study numerically how multiple deformable capsules squeeze into a constriction.
This situation is largely encountered in microfluidic chips designed to manipulate living cells, which are soft entities.
We use fully three-dimensional simulations based on the lattice Boltzmann method to compute the flow of the suspending fluid, and on the immersed boundary method to achieve the two-way fluid-structure interaction.
The mechanics of the capsule membrane elasticity is computed with the finite element method.
We obtain two main states: continuous passage of the particles, and their blockage that leads to clogging the constriction.
The transition from one state to another is dictated by the ratio between the size of the capsules and the constriction width, and by the capsule membrane deformability.
This latter is found to enhance particle passage through narrower constrictions, where rigid particles with similar diameter are blocked and lead to clogging.
\end{abstract}
\date{\today}
\maketitle
\section{Introduction}
\begin{figure}[b]
\centering
\includegraphics*[width = 0.45\textwidth]{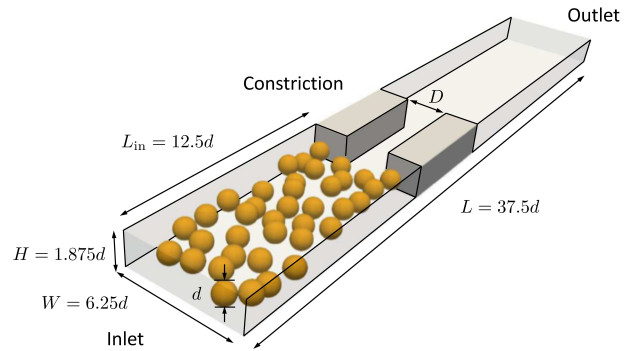}
\caption{The numerical setup used to study the flow of capsules (orange-colored spheres).
The microfluidic constriction forms an angle of $90$ degrees with the channel walls.
The flow direction is from the left-lower pre-constriction chamber to the right-upper post-constriction chamber.
The main geometrical parameter is the ratio of the channel width $D$ to the capsule diameter $d$: $D/d$.
$D$ is varied while holding $d$ constant.}
\label{fig:numerical_setup}
\end{figure}

The flow of particles is largely encountered in microfluidic devices designed to manipulate, sort, or characterize micro-sized artificial particles or living cells.
The performance of such devices can dramatically be hindered by the clogging events that take place at the entrance of channels, especially when the particle size is of the same order as the microchannels width or when multiple particles arrive suddenly at narrow passages \cite{Dressaire2017,Hong2017,vanZwieten2018,Zhang2018}.
While there is an increasing need to improve the capability of microfluidic devices to handle high particle throughput, this study is carried out to address the scenarios that emerge when multiple soft particles are pushed to flow through an abrupt 90 degrees microfluidic constriction, see Fig.~\ref{fig:numerical_setup}.
The constriction has a basic geometrical shape, but it is one of the most commonly encountered microfluidic devices engineered with the soft lithography technique. 
The flow of soft particles into constrictions is also encountered \textit{in vivo}.
For example, in blood vessels where the accumulation of fat on their walls may obstruct the flow of red blood cells and other particles, such as drug-carrier particles.

Recent experimental studies have investigated the clogging phenomenon by rigid particles \cite{Marin2018,Souzy2020}, which are found to form a stable arch at the entrance of the microfluidic constriction, and thus, block the channel. 
The neck-to-particle size ratio is found to be the leading parameter that determines the transition threshold to clogging, with permanent clogs systematically formed for a neck-to-particle size ratio below 3. For sufficiently large neck-to-particle size ratio, the particles flow either continuously or intermittently depending on the solid volume fraction. The corresponding dynamics has been characterized by stochastic mathematical models. In the case of soft particles, and to the best of our knowledge, the existing literature focuses solely on the flow and deformation of single isolated particles passing through constrictions \cite{Rorai2015,Kusters2014,LeGoff2017,Luo2017,Fai2017,Lei2019}.  
In this study, we extend these previous works and get one step toward mimicking living cells in microfluidic chips by considering a suspension of soft particles. 
\begin{figure*}
\centering
\includegraphics[width = 0.9\textwidth]{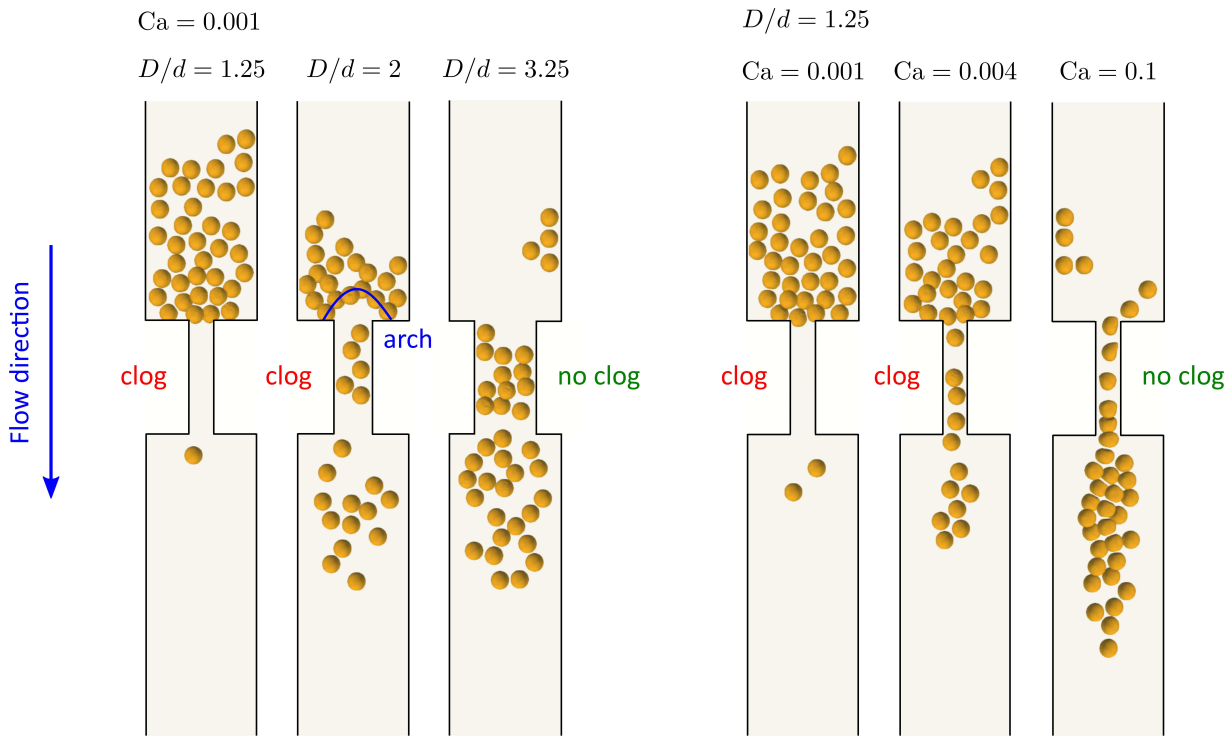}
\caption{Motion of multiple capsules ($N=38$) at a microfluidic constriction for a given flow strength.
The left panel shows the effect of the constriction aperture $D/d$, and the right panel the effect of particle deformability ${\rm Ca}$.
Easy passage of the particles is achieved at large $D/d$ and ${\rm Ca}$.
Arch structures form at the entrance of the constriction for narrower constrictions and rigid particles.
At large ${\rm Ca}$ and narrow constrictions, the particles cross the constriction one by one, while forming a regular train and exiting the constriction as a jet.}
\label{fig:clog_nonclog}
\end{figure*}

We perform fully three-dimensional simulations as used in Ref.~\cite{Kusters2014} to study the collective motion of multiple particles through a constriction, while varying two control parameters: i - the particle deformability to cover the range of rigid (non-deformable) and soft (deformable) particles, and ii - the width of the constriction.
The details of the numerical setup are shown in Fig.~\ref{fig:numerical_setup}.
The deformability is found to affect the transition to clogging, and serves as an additional key parameter in predicting the mechanism and the behavior of particles squeezing into a microfluidic constriction.
We report a state-diagram that depicts whether, or not, particles clog the channel depending on their deformability and the ratio of the constriction width to the capsule diameter.
Moreover, we characterize the particle passage by analyzing the dynamics of the number of particles passing the constriction.
\section{Setup and method}
We consider a microfluidic channel with dimensions $L = 37.5 d$ (length), $W = 6.25 d$ (width) and $H = 1.875 d$ (height) as shown in Fig.~\ref{fig:numerical_setup}, where $d$ is the diameter of the capsules at rest when they adopt a spherical equilibrium shape.
We set $d=2r=8$.
All units are given in lattice units as defined by the lattice Boltzmann method used to compute the flow.
The technical details about the numerical methods used to obtain the results are given in the Appendix.
Both the suspending and the encapsulated fluids are considered to have identical kinematic viscosity $\nu$ and mass density $\rho$. 
The channel inlet contains initially $38$ non-Brownian neutrally buoyant particles with no viscosity constrast.
All the particles have the same geometrical and mechanical properties (i.e. a monodisperse suspension).
They are randomly placed at the entrance region of the constriction. The present work is limited to fixed number of particles, and how this parameter alters the behavior of the system is left for a future study.

A constriction of length $L_{\rm obst} = 6.25 d$ and width $D$ is placed at a distance $L_{\rm in} = 12.5 d$ from the inlet of the channel having the width $W$.
A body force ${f}_z$ is applied in the \textit{z}-direction to generate a flow with a parabolic velocity profile, whose mid-plane velocity $u_{\rm max}$ in absence of particles and the constriction is given by
\begin{equation}
{u}_{\rm max} = \frac{{f}_z\,{W}^2}{8\rho\nu}.
\end{equation}
Here, $f_z = 2.22\times 10^{-6}$ is set in all simulations.

We study the influence of two dimensionless control parameters: i - the aperture defined as ratio of the constriction width to the capsule diameter $D/d$, and ii - the particle deformability quantified by the capillary number,
\begin{equation}
\mathrm{Ca} = \frac{\rho\nu r \dot\gamma}{\kappa _{\rm s}}, 
\end{equation}
where $\dot{\gamma} = {4 u_{\rm max}}/{W}$ is the measured shear rate at the wall.
We use also ${\rm K} = 1/{\rm Ca}$ that expresses the dimensionless elastic modulus of the membrane, and that characterizes the rigidity of the particles.
$D/d$ is set to desired values by varying $D$, while keeping $d$ constant. 
The value of ${\rm Ca}$ is varied by varying only the shear elastic modulus $\kappa_{\rm s}$, while holding all the other parameters constant.
\section{Results}
\subsection{Clog and no clog states}
\label{sec:states}
All the simulations are carried out on $40$ CPUs during $60$ hours at the Reynolds number ${\rm Re} = u_{\rm max}R/\nu = 0.1$, as is encountered in microfluidic flows. 
Snapshots showing both the effect of varying the constriction aperture $D/d$ (left panel) and the effect of varying the particle deformability ${\rm Ca}$ (right panel) are shown in Fig.~\ref{fig:clog_nonclog}.
The aperture degree is varied from $1.25$ (left snapshot) to $D/d = 3.25$ (right snapshot). 
The capillary number is set to $\mathrm{Ca} = 0.001$ to model rigid particles
because when $\rm{Ca} \ll 1$ the capsules deform weakly, and thus, behave mechanically as rigid particles, see Ref.~\cite{SM1}. 
For the case of $D/d = 1.25$ (narrowest constriction), two particles are found to be sufficient to clog permanently the constriction entrance.
For $D/d = 2$, more than two particles clog the constriction by building a stable arch, while the fluid still continues to flow.
For $D/d = 3.25$ (widest constriction), the aperture is large enough to allow and maintain a continuous flux of particles without observing any clog formation.
In the right panel of Fig.~\ref{fig:clog_nonclog}, we hold the same aperture degree $D/d = 1.25$, and we vary only the capillary number. 
For non-deformable particles, with $\mathrm{Ca} = 0.001$, again only two particles are sufficient to clog the constriction. 
When increasing ${\rm Ca}$, the particles deform further, and thus, they can squeeze easily through the constriction, see Ref.~\cite{SM2}. 
No clog event is observed for particles with large deformation capability under flow.
At large ${\rm Ca}$ and narrow constrictions, the particles cross the constriction one by one, while forming a regular train and exiting the constriction as a jet.
The reported results are observed independently of the initial random positions of the particles.
\subsection{State-diagram}
\label{sec:state-diagram}
For a given number of particles ($N=38$) and a given applied flow strength ($f_z = 2.22\times 10^{-6}$), we explore the state-diagram that gives the blockage status as a function of the constriction aperture $D/d$ and the capillary number ${\rm Ca}$.
This latter is taken smaller than $0.1$ to avoid the limit beyond which the capsules, whose membrane elasticity follows a Neo-Hookean law, are susceptible to undergo continuous elongation \cite{Hu2013}.

\begin{figure}
\centering
\includegraphics*[width = 0.49\textwidth, angle = 0]{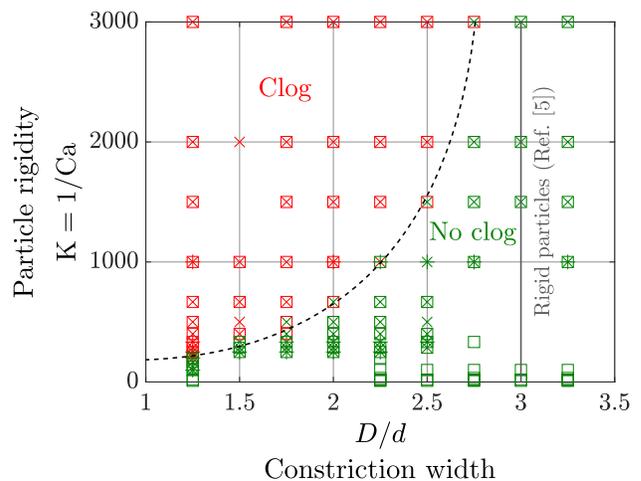}    
\caption{The clog and the no clog states of multiple capsules ($N=38$) obtained when varying their deformability ${\rm Ca}$ (or rigidity ${\rm K}$) and the constriction width $D/d$, while holding the same applied flow strength. The deformability character of the particles reduces the threshold of the transition to the no clog state below $D/d=3$ measured experimentally for rigid particles \cite{Marin2018}.
Simulations with three initial conditions are represented with different symbols: hollow squares, saltires and crosses.
}
\label{fig:state_diagram}
\end{figure}

Figure \ref{fig:state_diagram} reports the state-diagram, where we distinguish clearly two main regions representing the clog (red symbols) and the no clog states (green symbols).
The border separating the two regions (dashed black line) depends on the deformability of the capsules and their aspect ratio with respect to the constriction width.
It is also sensitive to the initial spatial arrangement of the particles.
Simulations with three initial conditions are represented with different symbols: hollow squares, saltires and crosses.

In the limit of weakly deformable capsules, i.e. when ${\rm Ca} \rightarrow 0$ (${\rm K} \rightarrow \infty$), the transition from clog to no clog states occurs at a critical value that approaches $D/d=3$, which is measured experimentally for rigid spherical particles and when using a converging-diverging constriction \cite{Marin2018}.
By choosing a basic step-like shaped constriction here, we intentionally rule out the angle of the constriction to appear as an additional parameter \cite{Lopez-Rodriguez2019}.
Capsules with a large degree of deformability, beyond the threshold $\rm{Ca} = 0.005$ (below $\rm{K} = 200$), can pass without clogging the constriction independently of the aspect ratio, even for particles having almost the same size as the constriction width ($D/d \rightarrow 1$).

Here, we report the effect of the deformability of multiple fluid-filled particles (not matrix microgel particles).
Soft particles can cross the constriction easily, and by varying their elasticity the threshold for the transition to the no clog state drops down.
This border is not sharp since it is sensitive to the initial arrangement of the particles.
There, both the clog and the no clog states may emerge.
In contrast, far from this blurry border, only one of the two states emerges with $100\%$ probability.
In our previous work Ref.~\cite{Kusters2014}, the state-diagram has been reported only for a single deformable particle for which a single particle can always cross the constriction when $D \geq d$.
Moreover, other works dealing with soft particles (e.g. Ref.~\cite{Harth2020}) have not considered varying the deformability as a control parameter.
\subsection{Dynamics of the particle passage}
\label{sec:escapees}
\begin{figure}
\centering
\subfloat[]{\includegraphics[width = 0.4\textwidth, angle = 0]{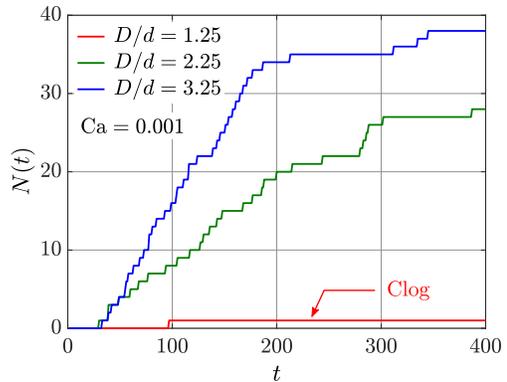}}
\\
\subfloat[]{\includegraphics[width = 0.4\textwidth, angle = 0]{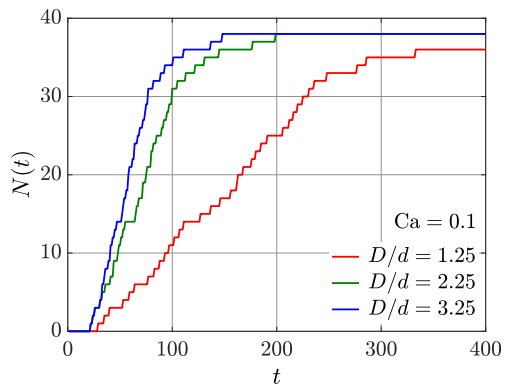}}    
\caption{Dynamics of the number of escapees through the constriction $N(t)$ for rigid spherical particles with ${\rm Ca}=0.001$ (upper panel) and for soft particles with ${\rm Ca}=0.1$ (lower panel) at various constriction widths $D/d$.
$N(t)$ increases approximately linearly faster, then it slows down non-linearly before it adopts a plateau, whose value corresponds to the total number of particles $N$ for the no clog state and to a smaller value $N(\infty) < N$ for the clog state.
No clogging event takes place for deformable particles, and the evacuation process is fast.}
\label{fig:nb_escapees}
\end{figure}

The number of escapees $N(t)$, defined as the number of particles that have passed through the constriction at time $t$, is given in Fig.~\ref{fig:nb_escapees} for stiff particles with $\rm{Ca} = 0.001$ (upper panel) and for soft particles with $\rm{Ca} = 0.1$ (lower panel). 

Three constriction aperture degrees are considered in each figure: $D/d = 1.25$, $2.25$, and $3.25$.
For stiff particles and at small aperture $D/d = 1.25$, a clog forms, as shown in Fig.~\ref{fig:clog_nonclog}. 
At $D/d = 2.25$ and $3.25$, the number of escapees over time $N(t)$ evolves in the same way linearly at the early stage and then non-linearly.
This means the flux of the particles across the constriction $dN(t)/dt$ is almost constant at the beginning of the simulations.
It adopts a greater slope for $D/d = 3.25$ for which three particles can pass at the same time through the constriction, while only two particles can pass for $D/d = 2.25$. 
In the second stage the number of escapees increases non-linearly and slowly. 
This latter represents the passage of the remaining particles that were trapped in the corners close to the channel walls that are located just at the constriction entrance, where the flow speed is lower.
Moreover, these particles need to overcome the step of the inlet constriction, which further increases the time needed for them to pass through the constriction.

\begin{figure}
\centering
\subfloat[]{\includegraphics[width = 0.4\textwidth]{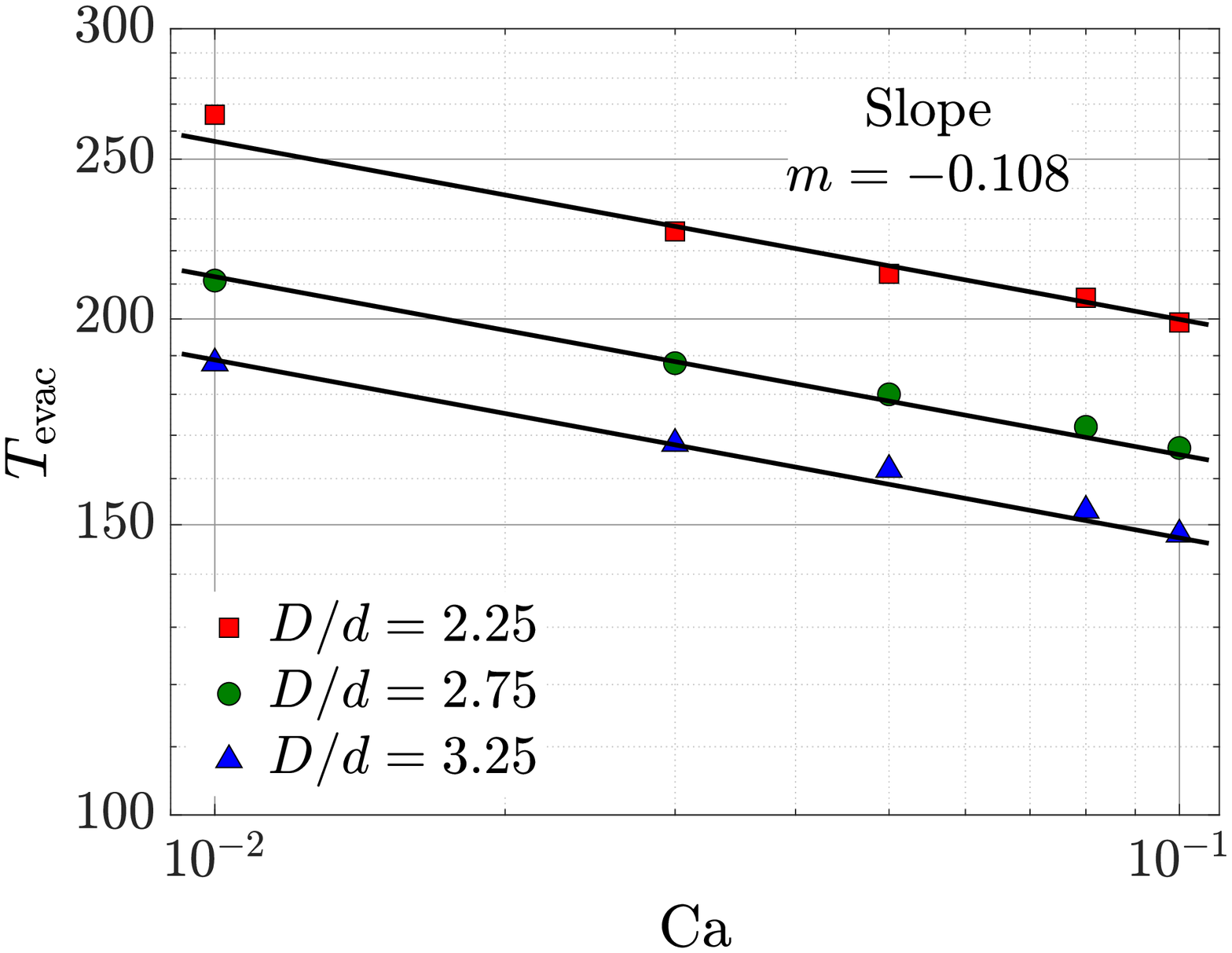}}
\\
\subfloat[]{\includegraphics[width = 0.4\textwidth]{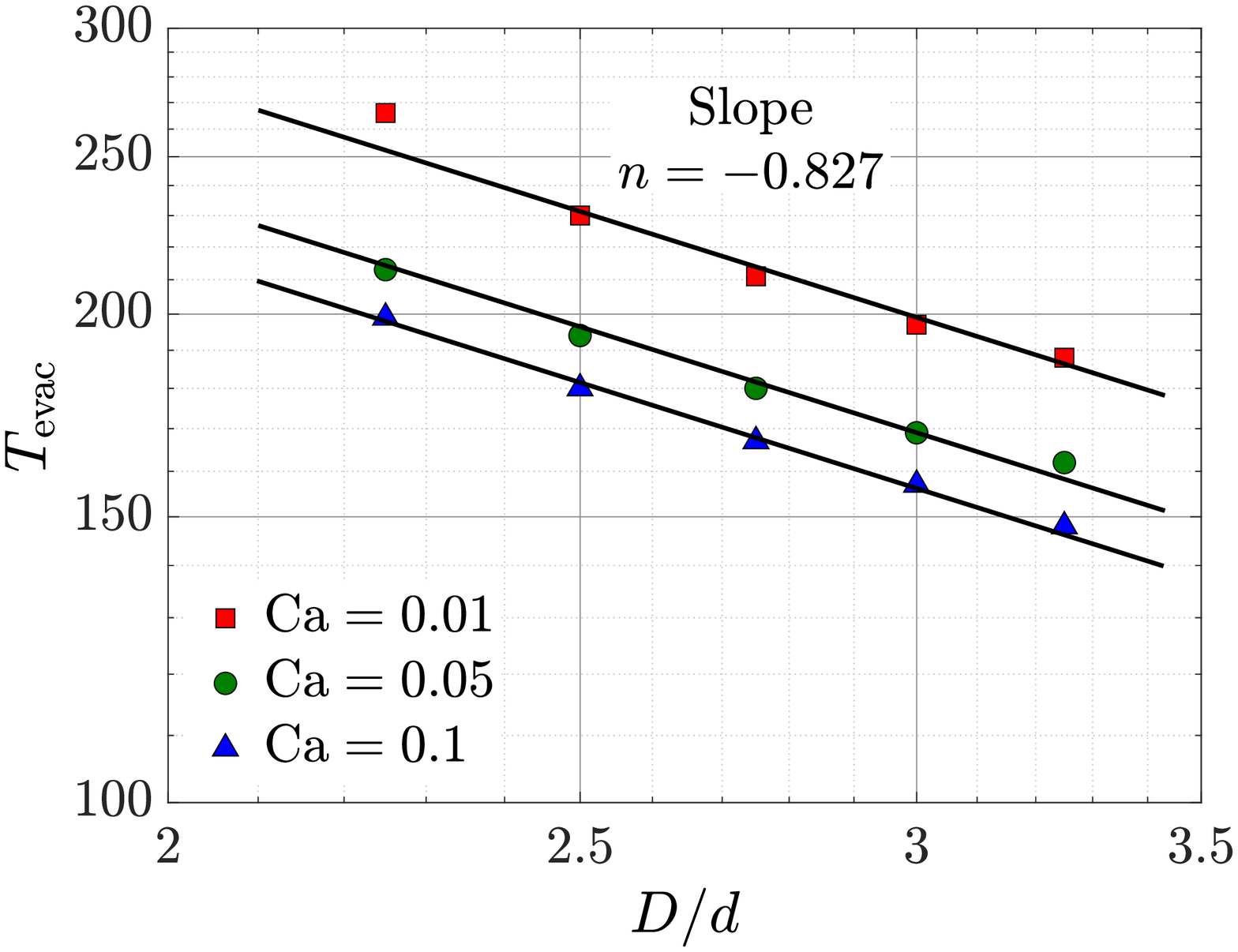}}
\caption{Evacuation time $T_{\rm evac}$ needed for $38$ capsules to pass the constriction as a function of the aperture $D/d$ (the upper panel) and the capillary number ${\rm Ca}$ (the lower panel).
$T_{\rm evac}$ is a decreasing non-linear function of both ${\rm Ca}$ and $D/d$.}
\label{fig:evacuation_time}
\end{figure}

When the particles are highly deformable, as shown in the lower panel of Fig.~\ref{fig:nb_escapees} for the case of $\rm{Ca} = 0.1$, all the particles pass easily through the constriction, even for the narrower constriction of $D/d = 1.25$. 
One may notice a plateau at $38$ (which is the initial total number of particles contained in the channel inlet) for $D/d = 2.25$ and $3.25$. 
This corresponds to the scenario where all particles have successfully passed through the constriction, and none of them is left in the inlet compartment.
This is the total evacuation, to which we associate the evacuation time $T_{\rm evac}$ that measures the time needed to evacuate all the particles, and which is found to increase when narrowing the constriction width.
$T_{\rm evac}$ is similar to the time needed to empty totally one compartment of a sand-clock made of granular particles, while in this study the particles are soft and are suspended in a viscous fluid.
It is reported in Fig.~\ref{fig:evacuation_time} as a function of $D/d$ and ${\rm Ca}$.
$T_{\rm evac}$ clearly decreases as the aperture $D/d$ and the capillary number ${\rm Ca}$ increase because particle passage is easier when the constriction is wider and the particles are more deformable. 
The dependency of $T_{\rm evac}$ on both $D/d$ and ${\rm Ca}$ is non-linear. 
The derivation of a scaling law $T_{\rm evac}=f(D/d,{\rm Ca})$ would be practical in designing microfluidic constrictions with desired throughput of soft particles.
However, in absence of a theoretical model to guide the scaling, we were only able to extract the exponents that give the dependency of $T_{\rm evac}$ on $D/d$ and ${\rm Ca}$ for the range of parameters available to our simulations,
\begin{equation}
T_{\rm evac} \propto \left( \frac{d}{D}\right)^{0.827}\quad \text{and} \quad T_{\rm evac} \propto {\rm Ca}^{-0.108}
\end{equation}
using the data set plotted on log-log scale in both panels of Fig.~\ref{fig:evacuation_time}.
The exponent related to ${\rm Ca}$ is low, here, that means the constriction aperture $D/d$ is the leading parameter.
However, its exponent of the proportionality $0.827$ is larger than $1/2$ known for the evacuation time of granular dry particles through a pore due to the contribution of the particle deformation and the presence of a suspending fluid in the present study.
\section{Discussion and conclusions}
\label{sec:conc}
Multiple deformable capsules passing through a microfluidic constriction exhibit similar dynamics as observed for a single capsule reported in Ref.~\cite{Kusters2014}, with slight differences due to the collective motion that is rendered cooperative due to the particle deformation.
The capsules either pass the constriction or they get stuck by building an arch at the constriction entrance.
Here, the flow is not blocked since the fluid can flow around and through the blocked capsules.
This study has examined both the effect of the constriction confinement and the particle deformability degree for a fixed number of particles and for a given applied flow strength.
The suspension is dilute when compared to the overall volume of the channel and the constriction, but it becomes dense at the constriction's entrance.
For the same flow strength, the transition from the blockage state to the passage is controlled by the size aspect ratio and the capsule deformability.
Capsules with extremely weak deformability show an almost similar threshold, as measured experimentally by Marin \textit{et al.} \cite{Marin2018} for rigid particles $D/d = 3$.
The capsules have revealed the presence of two types of the blockage states: permanent and transient, which lead to a non-sharp border in the state-diagram.
The dynamics of the number of particles crossing the constriction evolves linearly in the beginning of all the simulations, then non-linearly before it adopts a plateau when all the particles have been evacuated.
The complete passage of the capsule through the constriction is carefully analyzed, and it may be used when designing microfluidic devices for handling high throughput of soft particle suspensions.
At large ${\rm Ca}$ and small aperture ($D/d \rightarrow 1$), the particles cross the constriction one by one by forming a regular train, which could be exploited for diagnosis in the case of living cells.
This study has examined the role of both the constriction confinement and the particle deformability for a fixed number of particles, while further exploration of the parameter space is left for a future work, where we plan to vary the geometry of the constriction and the number of particles.
The small number of particles used in the present study does not lead to strong stochastic behavior as observed for a large number of particles.
For example, when flowing through a silo \cite{Harth2020}, which is described by statistics and probabilistic mathematical models.
Here, faraway from the border of the state-diagram only one of the clog and no clog states occurs.
At the vicinity of the border these two states may emerge with a probability due to the weak, but non-negligible effect of the initial positions of the particles.
\section*{Appendix: Numerical method}
\subsection{Fluid flow solver}
We use the lattice Boltzmann method (LBM) as a mesoscopic numerical method which allows to recover solutions of the Navier-Stokes equations.
We introduce very briefly the method, while the readers are invited to check out existing textbooks for more details \cite{Succi2001,Kruger2016}.
The spirit of the LBM consists of streaming a distribution function $f_i$ that gives the probability to find a number of fluid particles on a lattice node $\mathbf{r}$, at a discrete time step $t$, and with a discrete velocity $\mathbf{e}_i$.
The dynamics of $f_i$ is given by
\begin{equation}
f_i(\mathbf{r}+\mathbf{e}_i, t + 1) - f_i(\mathbf{r},t) = \Omega_i(\mathbf{r},t) + F_i(\mathbf{r},t).
\label{eq:lbe}
\end{equation}
$\Omega_i(r,t)$ on the right hand side is the collision operator.
The time and space steps are both taken to be unity.
Here, we use the Bhatnagar-Gross-Krook (BGK) collision operator \cite{Bhatnagar1954},
$\Omega_i = -\frac{1}{\tau} [f_i(\mathbf{r},t)-f_i^{\rm eq}(\mathbf{r},t)]$, that expresses the relaxation of $f_i$ toward its equilibrium $f_i^{\rm eq}$ within the relaxation time $\tau$.
$f_i^{\rm eq}$ is given as a truncated expansion of the Maxwell-Boltzmann distribution for the velocities in an ideal gas. 
External applied forces, including the membrane forces, are incorporated in Eq.~\ref{eq:lbe} through the source term $F_i$ such that
\begin{equation}
F_i(\mathbf{r},t) = \omega_i \left(1 - \frac{1}{2\tau} \right) \left(\frac{\mathbf{e}_i-\mathbf{u}}{c_{\rm s}^2}+\frac{\mathbf{e}_i \cdot \mathbf{u}}{c_{\rm s}^4}\mathbf{e}_i\right)\cdot\mathbf{F}(\mathbf{r},t),
\end{equation}
where $\mathbf{F}(\mathbf{r},t)$ accounts for either the body force or the membrane forces.
$c_{\rm s} = 1/\sqrt{3}$ is the lattice speed of sound and $\omega_i$ are the lattice weights which, for the three-dimensional lattice with 19 velocities (D3Q19) used here, read as $1/3$, $1/18$ and $1/36$ for $i=1$, $i=2\dots7$, and $i=8\dots19$, respectively.
We impose no-slip boundary conditions on the channel walls using mid-grid bounce-back boundary conditions.
\subsection{Mechanics of particle deformation}
We consider strain-softening capsules with zero-thickness membranes that exhibit Neo-Hookean 2D hyperelastic mechanical behavior with the energy \cite{Barthes-Biesel2016}
\begin{equation}
E_{\rm s} = \frac{\kappa _{\rm s}}{2} \int_A \left[I_1 - 1 + \frac{1}{I_2 + 1}\right] {\rm d}A,
\label{eq:strain_energy}
\end{equation}
where $\kappa _{\rm s}$ is the shear elastic modulus, $I_1=\lambda_1^2 + \lambda_2^2 -2$ and $I_2 = \lambda_1^2 \lambda_2^2 -1$ are the two deformation invariants, $\lambda_1$ and $\lambda_2$ are the principal stretching ratios, $A$ is the surface of each capsule, and ${\rm d}A$ the surface element.
In addition, we enforce the constraint of the capsule volume conservation by using an energy that gives the cost of any deviation of the actual volume of the capsule $V$ from its original value $V_0$ that is the volume of the spherical capsule at rest, 
\begin{equation}
E_{\rm v} = \frac{\kappa_{\rm v}}{2} \frac{(V-V_0)^2}{V_0},
\label{eq:volume_conservation}
\end{equation}
where $\kappa_{\rm v}$ is a numerical parameter whose value is set large enough to fullfill the volume conservation constraint.
Non-physical wrinkles may also emerge at the surface of the capsules, and are consequently avoided by applying a bending force $\mathbf{F}_{\rm b}$, which is derived as a functional derivative of the Helfrich energy originally proposed for lipid membranes \cite{Helfrich1973},
\begin{equation}
\mathbf{F}_{\rm b}(\mathbf{x}_i) = 2 \kappa_{\rm b} [2H (H^2 - K) + \Delta_{\rm s} H]\mathbf{n},
\end{equation}
where $\kappa_{\rm b}$ is the bending modulus, $H = \frac{1}{2}\sum_{i=1}^2 c_i$ is the mean curvature, $K=\prod_{i=1}^2 c_i$ is the Gaussian curvature, $c_i$ is the principal curvature, $\Delta_{\rm s}$ is the Laplace-Beltrami operator and $\mathbf{n}$ the normal vector pointing outward from the membrane. 
The bending modulus is chosen such as the dimensionless number $B=\kappa_{\rm b}/(\kappa_ {\rm s} r^2)$, that quantifies the relative importance of the bending rigidity with respect to the shear elasticity, is small.
In this way the bending force mitigates wrinkle formation, while it does not influence globally the dynamics and deformation of the capsules. 
$H$, $K$ and $\Delta_{\rm s}$ are computed following a discrete differential geometry operators approach \cite{Guckenberger2017}.  

The membrane of each capsule is discretized into $1280$ triangular elements, and the force on each membrane node $\mathbf{x}_i$, with $i$ refering to the index of the node, is evaluated following the principle of virtual work such that
\begin{equation}
\mathbf{F}_{\alpha}(\mathbf{x}_i) = -\frac{\partial E_\alpha}{\partial \mathbf{x}_i}.
\label{eq:virtual_work}
\end{equation}
The subscript $\alpha$ denotes either $\lbrace {\rm s} \rbrace$ for the strain energy, $\lbrace {\rm v} \rbrace$ for the volume energy or $\lbrace {\rm b} \rbrace$ for the bending energy.
The derivatives needed to evaluate the membrane forces are computed numerically using the finite element method Ref.~\cite{Kruger2012}.
A short-range repulsive force is implemented to mimic the hydrodynamic lubrication force, and to avoid overlap between particles or between particles and walls,
\begin{equation}
\mathbf{F}_{rep} = \left\{
\begin{array}{rcr}
&\bar{\epsilon}[(\frac{1}{d_{ij}})^2 - (\frac{1}{\delta_0})^2] \frac{\mathbf{d}_{ij}}{d_{ij}} \quad \text{if} \quad d_{ij} < \delta_0 \\
&\mathbf{0} \quad \text{if} \quad d_{ij} \ge \delta_0
\end{array}
\right. ,
\label{eq:frep}
\end{equation} 
with $\bar{\epsilon}$ being the strength of the force, and $d_{ij}$ is the surface-to-surface distance between particles $i$ and $j$ or the distance between particle $i$ and the solid node $j$ on the wall. The repulsive force vanishes when $d_{ij}$ is larger than the cutoff distance $\delta_0 = 1$.
\subsection{Fluid-structure interaction}
The two-way coupling between the fluid flow and the capsule dynamics is realized using the immersed boundary method (IBM), which is a front-tracking method developed originally by Peskin to study blood flow in the heart \cite{Peskin1977}. 
The IBM consists of coupling a moving Lagrangian mesh $\delta \Omega$ representing the capsule membrane and a stationary Eulerian grid $\Omega$ \cite{Peskin2002}, where the flow is computed with the LBM.
The method has two main steps:

\textit{Advection - }The flow advects all the capsules' mesh nodes as if they are massless pointwise particles. 
Once the fluid velocity field ${\bf u}(x,y,z,t)$ is computed by the LBM on the Eulerian mesh, the velocity of each membrane node ${\bf u}(s_1,s_2,t)$, where $(s_1,s_2)$ are curvilinear coordinates, is estimated by interpolation of the velocities of its neighboring fluid nodes using a function $\delta$,
\begin{equation}
{\bf u}(s_1,s_2,t) = \int _{\Omega} \delta({\bf r}(x,y,z,t),{\bf r}(s_1,s_2,t)) {\bf u}(x,y,z,t) {\rm d}{\bf r},
\end{equation}
with ${\bf r}(x,y,z,t) \in \Omega$, ${\bf r}(s_1,s_2,t) \in \delta \Omega$, and 
\begin{equation}
\delta({\bf r}_1,{\bf r}_2) = \phi(x_1,x_2)\phi(y_1,y_2)\phi(z_1,z_2),
\end{equation}
where
\begin{equation}
\phi(x_1,x_2) = \frac{1}{4}\left(1 + \cos \frac{\pi (x_ 1- x_2)}{2} \right)
\end{equation}
if $\vert x_1 - x_2\vert \leq 2$, $\vert y_1 - y_2\vert \leq 2$ and $\vert z_1 - z_2\vert \leq 2$, otherwise $\delta({\bf r}_1,{\bf r}_2) = 0$.
All the membrane nodes are then advected using the explicit Euler scheme:
\begin{equation}
{\bf r}(s_1,s_2,t+1) = {\bf r}(s_1,s_2,t) + {\bf u}(s_1,s_2,t) 
\end{equation}

\textit{Reaction - } 
When all the membrane nodes are advected, the overall capsule deforms into a new shape that is not necessarily its equilibrium shape, and thus, it tries to relax back to its lowest energy configuration.
By doing so it exerts a force back upon its surrounding fluid.
The forces exerted by the membrane in the Lagrangian mesh ${\bf F}(s_1,s_2,t)$ are computed with the finite difference method and are extrapolated to the fluid nodes using $\delta$ again as a weight in order to have a force field in the Eulerian grid ${\bf F}(x,y,z,t)$,
\begin{equation}
{\bf F}(x,y,z,t) = \int _{\delta \Omega} \delta ( {\bf r}(x,y,z,t),{\bf r}(s_1,s_2,t)) {\bf F}(s_1,s_2,t){\rm d}A.
\end{equation}
This force is plugged into the right-hand side of the LBM equation.
\newline
\section*{Acknowledgements}
The authors acknowledge financial support by the Deutsche Forschungsgemeinschaft (DFG) within the research unit FOR2688 ‘Instabilities, Bifurcations and Migration in Pulsatile Flows’ (grant number HA4382/8-1). 
CB and BK acknowledge the Ministère de l’Enseignement Supérieur, de la Recherche et de l’Innovation (MESRI) and the Biomechanics and Bioengineering Laboratory (BMBI) for financial support.
%
%

%
\end{document}